# Radio path length correction using water vapour radiometry

R.J. Sault, G.J. Carrad, P.J. Hall
CSIRO Australia Telescope National Facility, P.O. Box 76, Epping, NSW, 1710
(rsault@atnf.csiro.au; gcarrad@atnf.csiro.au; phall@atnf.csiro.au).

J. Crofts
Astrowave Pty Ltd.

Path length changes through the atmosphere cause significant errors for astronomical radio interferometry at high frequencies (e.g. 100 GHz). Path length differences typically give rise to a differential excess path of 1mm for antennas separated by 1km, and have fluctuation time scales of greater than 10 seconds. To measure these fluctuations, we are building a four-channel radiometer centred on the 22 GHz water line. By sensing the water vapour emission, the excess path can be deduced and corrected. Multiple channels give us robustness against various systematic errors, but gain stability of the radiometer of 1 pair in $10^4$ is still required.

**Introduction**

Astronomical radio interferometry is the practice of correlating the outputs from multiple antenna pairs to synthesise an aperture equivalent to the distances between the antennas. This enables much higher resolution to be achieved than that possible with a single dish radio telescope, as well as avoiding a number of systematic errors that plague single dish astronomy. Radio interferometry is routinely used to form images with resolutions of an arcsecond or better at wavelengths varying from meters to sub-millimeter. A detailed description of radio interferometry can be found in [1]. The Australia Telescope Compact Array (ATCA) is a radio interferometer operated by CSIRO near the NSW township of Narrabri. This array consists of 6 antennas on a rail track. Currently this array observes at wavelengths of 20 to 3cm, with a maximum baseline of 6 km. However the array is now being upgraded to operate at 12 and 3mm wavelengths.

Effectively, an interferometer array images by detecting the difference in arrival time of wavefronts between two antennas. To image to high resolution, the relative geometry of the array needs to be known with a precision of a small fraction of a wavelength. One of the major error sources in interferometry at short wavelengths is the difference in electrical length of the path through the atmosphere above different antennas. This is caused by variations in the refractive index of the atmosphere above different antennas.

**Refractive effects**

These variations distort the signal's wavefront, and result in the wave arriving at different antennas at slightly different times from those predicted. That is, there is an excess path difference resulting from these refractive index variations. These path differences must be accounted for during the processing of the data. At radio frequencies, the variations are almost entirely a result of 'parcels' of water vapour. Figure 1 gives a cartoon of the situation. These parcels come in all sizes, with the rms atmospheric path difference showing a power-law distribution with antenna separation (the so-called Kolmogorov distribution [2]). For Narrabri, a typical path difference resulting from atmospheric irregularities is 1mm for antennas 1km apart, and the power law is approximately a direct proportionality. A popular model is that the parcels of vapour are "frozen" into the wind flow, and so the path difference varies in the time needed for a parcel of water vapour to blow past an antenna. This is of order tens of seconds for the Narrabri array. If uncorrected, these fluctuations will completely prevent imaging when the rms fluctuations in path exceed about a sixth of a wavelength. Thus, when observing at a wavelength of 3 mm, we need to correct atmospheric path length changes to an accuracy of at least 50 μm. Radio astronomers have traditionally coped with atmospheric correction using two main approaches:

- Use periodic observations of strong nearby point sources to measure the atmospheric path difference. This becomes increasingly difficult at short wavelengths, where strong nearby sources are few, and the rate at which the atmospheric path length changes in both time and angle on the sky becomes too great.

- Use the source being observed to "self-calibrate" itself. This seemingly magical technique is possible when the number of measurements (which goes as the square of the number of antennas) greatly exceeds the number of unknowns (the path irregularities above each antenna). It is only possible provided the source is detectable on an interferometer baseline (with a signal-to-noise ratio of at least a few) in a period over which the atmosphere can be assumed to remain constant. Self-calibration is less effective at high frequencies, where sources tend to be weaker, receivers tend to be noisier, and the atmospheric time scale is shorter.

The atmosphere (like the weather!) is very variable. Best observing conditions are on clear winter nights. This is when the water content in the atmosphere is a minimum, and when there are no convection cells resulting from solar heating. Conversely summer afternoons generally experience the poorest observing conditions. The best sites for high frequency observing will be on high, dry areas, such as the Andes in Chile or the Tibetan plateau.

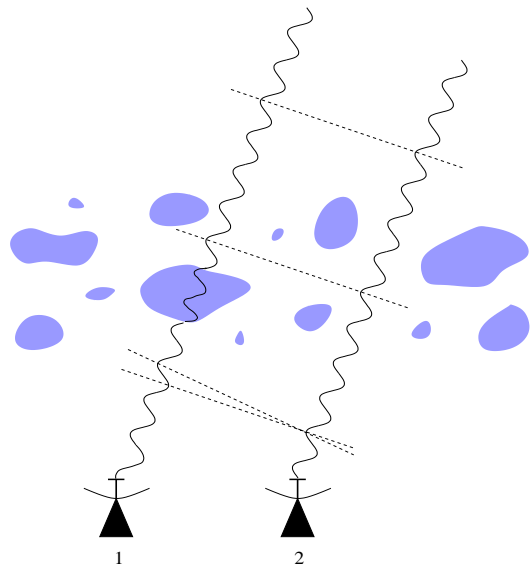

Figure 1: A cartoon of the effect of water on an astronomical signal. The plane waves of the signal are distorted by the presence of parcels of water vapour (diagram from M. Wiedner's thesis [5]).

**Water Vapour Radiometry**
Water vapour radiometry approaches path length estimation from a different perspective to that of traditional astronomical calibration. As the water vapour in the atmosphere radiates energy at centimetre wavelengths, we can deduce the water vapour fluctuations by monitoring the variation in sky brightness temperature. Figure 2 gives the sky brightness temperature expected at Narrabri as a function of frequency, and for different levels of atmospheric water content (generally expressed as the precipitable water vapour along the line of sight). Although the principle of water vapour radiometry has been appreciated for many years (e.g. [3]), it has found limited practical use in interferometry. Systematic errors and lack of the required gain stability have rendered the technique of limited utility.

A major issue with water vapour radiometry approaches is to differentiate between apparent changes in sky brightness caused by effects other than water vapour. These effects include changes in receiver noise, changes in ground pick-up as an antenna tracks, and contributions to brightness temperature from liquid water (e.g. clouds). In moderate to clear observing conditions, liquid water does not contribute significantly to path differences. However as liquid water radiates much more effectively than water vapour, the presence of small amounts of liquid water can still change the antenna temperature significantly.

To distinguish between water vapour and other effects, it is attractive to use a channelised radiometer that observes near a water line. By using multiple frequency measurements along the known shape of a water line, it is possible to distinguish between changes caused by water vapour and various systematic errors. That is, near a water line, the spectrum of water vapour is very different from that of other contributions to antenna temperature. The basic idea is to estimate the excess path resulting from water vapour as

$$L_{vap} = a_1 T_1 + a_2 T_2 + a_3 T_3 \Lambda$$

where $T_1$, $T_2$, $T_3$, etc, are the antenna temperatures measured at several frequencies and $a_1$, $a_2$, $a_3$, etc, are weights used to combine these measurements. As seen in Figure 2, there are two water lines in the frequency range shown: at 22 and 186 GHz (there are also oxygen lines near 60 and 118 GHz). Pioneering work by Woody [4] was responsible for a reawakening in interest in this approach. This approach became more feasible because of the availability of better quality microwave components (more stable and wider bandwidth) and an appreciation of the importance of eliminating systematic errors.

It is interesting to note that the requirements for a water vapour radiometry system for an interferometer are both more stringent and more relaxed than a general system to measure atmospheric water content. The requirements are more stringent in that the path difference that must be measured is small (e.g. at least 50 μm). However a key difference to a general path measuring system is that only the *difference* in path is needed. The absolute electrical path resulting from water vapour, which is typically 50-300 mm at Narrabri, is of no interest. Because we are dealing with a compact array, the atmosphere above the antennas is largely identical. It is only the smaller, turbulent component of the water vapour (which is typically only 1% of the whole) which is of interest. Because the atmosphere above the different antennas is largely the same, the errors introduced by relating the measurements to models of the atmosphere are largely eliminated. Indeed, while models of the atmospheric emission are needed to design the radiometer system (e.g. to pick the best frequencies), they are of limited utility in running the system. When it comes to practical use, a more empirical approach of relating the observables (the sky brightness temperature measurements) to the path differences is more important.

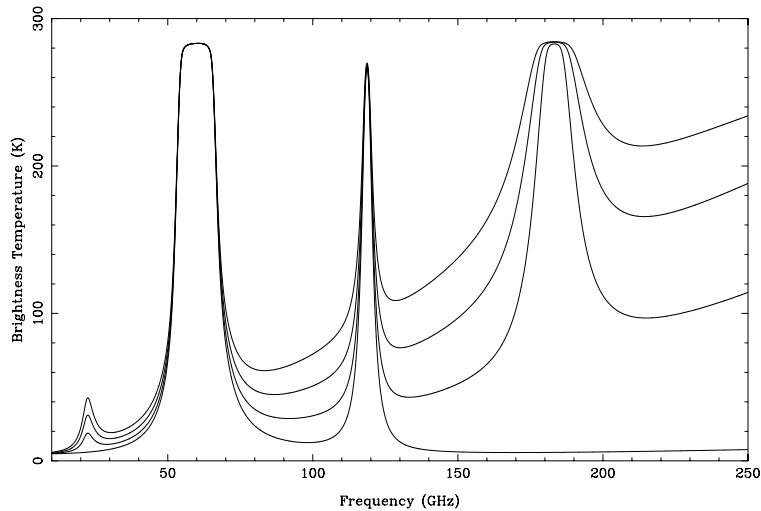

Figure 2: Sky brightness temperature as a function of frequency. The different curves correspond to different water contents in the atmosphere. There are water lines at 22 and 186 GHz, and prominent lines resulting from $O_2$ near 60 and 118 GHz.

**22 or 186 GHz?**
Although the water line at 186 GHz is a far stronger one (and so would seem more attractive), there are a number of arguments against this for a water vapour radiometry system for the ATCA:

- The ATCA antennas cannot operate at 186 GHz, and so a separate antenna would be required for the radiometer sensing if this frequency was to be used. However, so that they see the same cylinder through the atmosphere, it is desirable that the astronomical and water sensing antennas are identical. Thus, in addition to cost and complexity arguments, separate antennas are undesirable.

- The 186 GHz line saturates at the ATCA site (the atmosphere becomes opaque). To avoid saturation, a radiometer system would need to be offset from the water line. The resultant system would be very wide-band, and the sensitivity would be poorer than is suggested from simple arguments right at the 183 GHz line.

Although it is far weaker, the 22 GHz line is more attractive. This is reinforced by evaluating the sensitivity requirements: raw sensitivity is not a problem, even at 22 GHz. The success or failure of a water vapour radiometer approach is in the handling of gain stability and systematic errors.

**The Four Channel System**
Our design consists of a 4-channel radiometer near the 22 GHz line. To limit differences in systematics between the channels, they will all use the same amplifier. This places a limit on the spread in frequencies of these channels. Figure 3 shows the position and width of these four channels, as well as the expected change in brightness temperature for a change in path of 1mm. The many overlapping curves represent different meteorological conditions (the change in brightness is a function of temperature and pressure of the water vapour). The meteorological data for these are from radiosonde measurements taken over 10 years at Moree (a town near Narrabri, which has a similar meteorological environment). These curves are for radiosonde data which are believed to represent good weather for observing at millimeter wavelengths (i.e. the radiosonde data tend to represent clear winter nights). The frequencies to observe (and weights used to combine the measurements at the different frequencies) are optimised as follows:

- They are chosen to minimise error in the path estimate caused by receiver noise.
- They are such that the resultant estimate should be relatively insensitive to instrumental errors, ground pick-up and liquid water contributions.
- They should be relatively robust to different meteorological conditions.
- They should lead to an accurate determination of small changes in path (i.e. as is of interest in interferometry), rather than absolute path.

Our approach is to find those frequencies and weights that minimise the expected error in the path estimate, given projected receiver thermal noise and meteorological conditions. In doing this, we assume that the Moree radiosondes are representative of the atmosphere at Narrabri. The frequencies and weights are also chosen to be insensitive to offset, linear and quadratic changes in the antenna temperature with frequency, i.e. they rely on the water line being quite different from these in shape.

**Design Issues**

For the array operating at a wavelength of 3.5mm, the required accuracy in estimating the path is 50 μm. This corresponds to measuring the system temperature at 22 GHz to an accuracy of 20 mK, i.e. an accuracy in measuring antenna temperature of $1:10^4$. Measurements to a sensitivity of 20 mK are needed within the integration time of a few seconds. The system need not have long-term absolute stability: it is regularly checked by astronomically determining the path differences through the atmosphere by observing calibrator sources. Variations in the system on longer time scales than the astronomical calibration (typically tens of minutes) are not important.

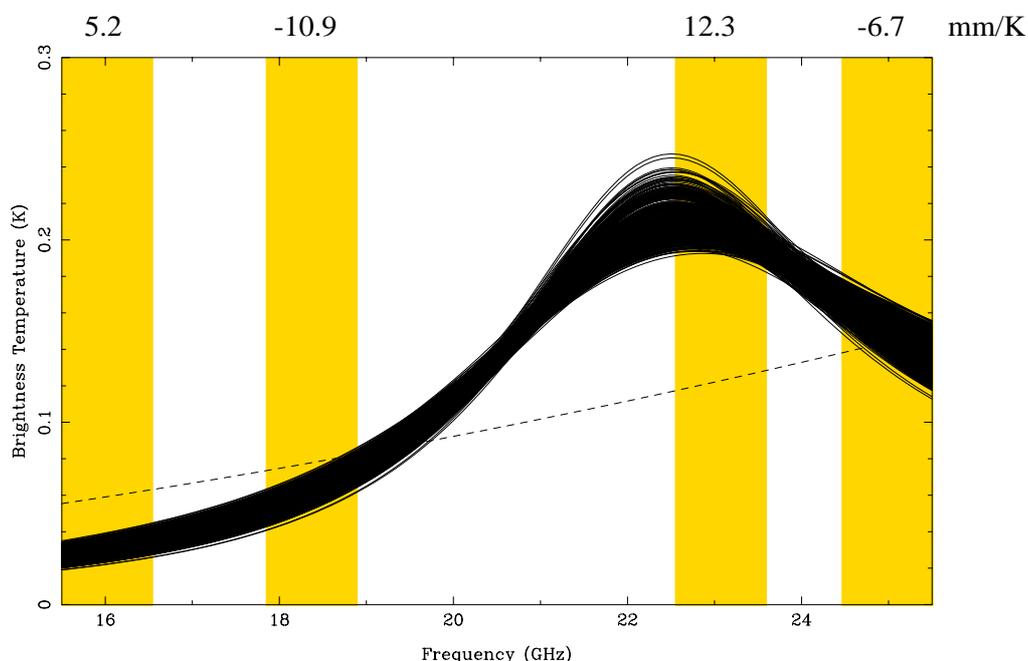

Figure 3: The grey bands indicate the four channels to be measured by the radiometer. The traces correspond to the change in sky brightness produced by 1 mm of excess path resulting from water vapour. The different traces are for different meteorological conditions. The dashed line gives the shape of the spectrum expected from liquid water.

Two prototype radiometers are under construction with amplifiers, signal splitter, band-pass filters and detectors on a common block under precision temperature control. This aims at eliminating thermally induced gain variations. The initial connectorised version will be upgraded to have components on a single substrate. A feed horn, to channel signals to the radiometer, will share space with other feeds of the high frequency upgrade of the ATCA. The radiometer and astronomy horns will be slightly offset: they will be looking though slightly different atmospheric columns. However this offset (about 5 arcminutes) is sufficiently small so that the cylinder of atmosphere that the two horns are sensitive to are almost completely coincident in the lower few kilometers of the atmosphere. The current design uses a separate 22 GHz receiver/horn to the astronomical 22 GHz package. Use of the 22 GHz feed and cooled astronomical receiver for the radiometer system is a possibility that is being considered. This would mean a simpler system and better system temperature for the radiometer. However, the prime requirements of astronomical and radiometer amplifiers (low noise and high stability respectively) are not necessarily compatible.